\documentclass{SVProc}
\usepackage{makeidx}  
\usepackage{verbatim}

\makeindex
\usepackage{url}

\begin{document}
\mainmatter   
\pagestyle{headings}  

\title{Easy and complex: new perspectives for metadata modeling using RDF-star and Named Graphs}
\titlerunning{Easy and complex: new perspectives for data modeling} 

\author{Florian Rupp \and Benjamin Schnabel \and Kai Eckert}
\authorrunning{Florian Rupp et al.} 

\institute{Stuttgart Media University, Stuttgart, Germany,\\
\email{rupp@hdm-stuttgart.de},\\
\texttt{http://wiss.iuk.hdm-stuttgart.de/}
}

\maketitle              

\begin{abstract}
The Resource Description Framework is well-established as a lingua franca for data modeling and is designed to integrate heterogeneous data at instance and schema level using statements.
While RDF is conceptually simple, data models nevertheless get complex, when complex data needs to be represented. Additional levels of indirection with intermediate resources instead of simple properties lead to higher barriers for prospective users of the data.
Based on three patterns, we argue that shifting information to a meta-level can not only be used to (1) provide provenance information, but can also help to (2) maintain backwards compatibility for existing models, and to (3) reduce the complexity of a data model.
There are, however, multiple ways in RDF to use a meta-level, i.e., to provide additional statements about statements.
With Named Graphs, there exists a well-established mechanism to describe groups of statements.
Since its inception, however, it has been hard to make statements about single statements.
With the introduction of RDF-star, a new way to provide data about single statements is now available.
We show that the combination of RDF-star and Named Graphs is a viable solution to express data on a meta-level and propose that this meta-level should be used as first class citizen in data modeling.

\end{abstract}

\section{Introduction}

Data can come in many different forms and there are nearly infinite ways to model data describing the same information. Depending on the use case, different forms might be preferable: a simple view on data is the easiest to work with and facilitates data reuse. On the other hand, more complex applications need advanced data models providing the full complexity of the data with all available details. The Resource Description Framework~(RDF)~\cite{noauthor_resource_nodate} provides a common basis for the publication of data in form of statements on the Web but makes no assumptions how these statements actually look like and how entities and their relations are used to represent the desired information. Thus, various modeling approaches can represent the same information. The Semantic Web aims at providing interlinked data based on the "follow your nose" principle~\cite{dodds_follow_2012}, i.e.: users and applications can get data about a resource by dereferencing its URI.

There are two possible solutions, when data is to be provided both in simple and complex\footnote{While it depends on the perspective, we consider data complex that uses several layers of properties and dependent classes to describe a resource.} forms for different applications: 

\begin{enumerate}
    \item Different data representations can be provided; in this case, a mechanism is required that allows the application to request one of these representations (see Section~\ref{related_work}). 
    \item The data is provided in a simple data model and additional information is provided on a meta-level in the form of statements that provide further explanations. 
\end{enumerate}

In this article, we discuss the latter approach. We argue that there are benefits of using such a meta-level approach versus the provision of different data formats: there is only one representation required, it is straight-forward to extend existing data representations while maintaining backwards compatibility and it fits the "follow your nose" principle of the Semantic Web where applications simply can ask if there is further information about a statement.

A meta-level can be created by providing a further statement describing a single or multiple statements. To provide statements about statements in RDF, a mechanism is needed to identify statements. The identification can happen in two ways: either a single statement is identified (statement-level) or a group of statements -- usually called a graph -- is identified and meta-statements are used to further describe all statements in this graph (graph-level).

For statement-level data there always has been the optional RDF reification, for graph-level statements, Named Graphs have been introduced. Particularly RDF reification, however, has two major shortcomings: it is inefficient for exchanging RDF data and writing queries to access statement-level metadata is cumbersome~\cite{hartig_foundations_2017}.

Now there is a new way to provide statement-level metadata within RDF itself with the recent specification of RDF-star\footnote{\url{https://www.w3.org/2021/12/rdf-star.html}}. With increasing support in RDF databases and tools and the specification of various serializations including TriG-star\footnote{Throughout this paper, we use TriG-star syntax in the examples. For reference this is a list of the implied namespaces, ex is the default or empty namespace: rdf: \url{http://www.w3.org/1999/02/22-rdf-syntax-ns#}; owl: \url{http://www.w3.org/2002/07/owl#}; ex: \url{http://example.org/}; dcat: \url{http://www.w3.org/ns/dcat#}; dct: \url{http://purl.org/dc/terms/}; foaf: \url{http://xmlns.com/foaf/0.1/}; prov: \url{http://www.w3.org/ns/prov#}; dbp: \url{http://dbpedia.org/property/}; dbo: \url{http://dbpedia.org/ontology/}; gn: \url{http://www.geonames.org/ontology#}; gndo: \url{http://d-nb.info/standards/elementset/gnd#}}, 
it provides all that is needed to use both statement-level metadata with RDF-star and graph-level metadata with Named Graphs. 

In this article, we explore the possibilities and limitations of RDF-star and Named Graphs with respect to the creation of data models which are simple and easy to use, but nevertheless contain the full complexity of the data on a meta-level so that it can be used when needed. We start with an overview on related work focusing on providing data to applications with different requirements. In Section~\ref{patterns}, we identify three abstract patterns that show how common use cases can be modeled with metadata. In Section~\ref{queries}, further examples about querying and working with the meta-level are given. The paper concludes with a discussion of the advantages and disadvantages of data modeling with a meta-level as first-class citizen.

\section{Related Work}
\label{related_work}

Different applications have different requirements towards data. Consider a personal blog that wants to provide a thumbnail view of a book versus a library that needs data about a book to include in its catalog. It is common practice in many open data portals that data can be accessed in different formats, e.g. Dublin Core for simple data vs. MARC for all the details. In the Semantic Web, however, this poses a problem. If a URI for a book is resolved automatically by an application, what data should be provided? In our paper this is not directly solved, but arguably our meta-level approach allows applications to decide if they want to access the additional information. For each resource there can only be one data description. The provider of the data needs to decide, how this data is modeled and which vocabularies are used. To address this shortcoming, the Dublin Core community developed the concept of \emph{Application Profiles}~\cite{heery_application_2000} to describe the data model for a specific application. Based on the notion of Application Profiles, there is a recent RFC draft to use HTTP for the negotiation of different data representations on content level, similar to the well-known content negotiation for different RDF serializations~\cite{svensson_indicating_2021}. Using such mechanisms, however, requires proper support in applications and it is not yet clear if this will actually be used in practice.

Another aspect is the provision of metadata to describe the actual data, mostly to provide information on the \emph{provenance} of the data \cite{orlandi_modelling_2011,eckert_provenance_2013,garijo_dublin_2013,perez_systematic_2018,sikos_provenance-aware_2020,orlandi_benchmarking_2021}. There are generally two aspects to consider: (1) how can the subject, i.e., the data, be identified to allow statements about it, and (2) how can the provenance of data be described. As we do not focus specifically on the description of provenance, we are mainly concerned with the former. 

Besides RDF-star~\cite{olaf_hartig_rdf-star_2021} and Named Graphs~\cite{carroll_named_2005}, other approaches have been proposed.
The oldest way to do this is \emph{RDF reification}, which is available since the very first versions of RDF, cf.~\cite{patrick_j_hayes_rdf_2014}. However, making statements about statements using reification is hard, mainly due to its verbose syntax and the fact that every statement needs to be reified, even when many statements share the same meta information. This might be a reason why reification is rarely used in real world applications\footnote{
To get evidence of reification usage in real world application we checked a given list of SPARQL endpoints. In favor we use the Wikidata SPARQL endpoint list at \url{https://www.wikidata.org/wiki/Wikidata:Lists/SPARQL_endpoints}. This examination was done automatically with a suitable SPARQL query. The list, however, contains 139 entries and can be, according to its small size, considered as a sample only. Merely 78 endpoints were reachable via GET or POST requests (on July 20th 2022). 7 out of 78 remaining endpoints were not suitable for this study, being no actual SPARQL endpoint or the domain has been sold. Out of 72 endpoints 5 are using reification. At just 7,7\% that's a small percentage. This supports the thesis that reification is rarely used.}

\emph{Singleton statements}~\cite{nguyen_dont_2014} use unique identifiers which are added to the predicate statement. Following this method, the meta-level can be expressed by using the predicate in combination with the assigned identifier. Singleton properties lead to compact representations of the meta-level, but obviously affect the querying of the data. There is support for singleton properties in some triple stores, which makes the application more feasible. Even then it is not recommended, for instance see the documentation of GraphDB\footnote{\url{https://graphdb.ontotext.com/documentation/free/devhub/rdf-sparql-star.html}, Singleton Properties.}).

These approaches all address statement-level metadata. There are many cases where an identification of single statements is not needed and even not helpful. Probably the best example is provenance information. Usually, many statements share the same provenance when they are created together by a single process.

For the separation of different descriptions of the same resource, as an alternative, \emph{Proxy entities}~\cite{isaac_antoine_europeana_2013} or resource maps in OAI-ORE~\cite{lagoze_ore_2008} have been proposed, that use proxies as distinct placeholder entities that all represent the same actual entity but with different URIs to be able to distinguish statements about the entity from different sources. This raises the complexity of the data and requires applications to correctly interpret the proxy entities. 

A problem with Named Graphs is that organizing structures, for example nested graphs or subgraphs can only be represented via additional data describing \emph{graph relationships}. This can be used to identify provenance information in hierarchical graphs~\cite{eckert_provenance_2013}. A proposal for the structured use of several named graphs are \emph{nano publications}~\cite{bucur_unified_2020} that also explicitly support provenance.

\section{Meta Modeling Patterns}
\label{patterns} 

Based on our experience with data applications, we can identify three abstract modeling patterns that illustrate the benefits of a meta-level. For each pattern, we formulate a problem statement, the solving approach, an example and a conclusion. 
The three patterns are:

\begin{itemize}
    \item Pattern 1: RDF provenance modeling in the meta-level.
    \item Pattern 2: Extending an existing data model.
    \item Pattern 3: Shifting proxy entities to the meta-level.
\end{itemize}

We will not distinguish statement-based and graph-based metadata. With RDF-star and Named Graphs, there are now viable solutions for both levels and it is best decided by the data modeler or the data provider which one is more suitable. This of course requires that consuming applications understand both approaches, i.e, TriG-star needs to be supported as serialization. 

\subsection{Pattern 1: RDF provenance modeling in the meta-level}
\label{pattern1}
\paragraph{Problem statement:}
The provision of provenance data \emph{in RDF} is straight forward, for example using PROV or Dublin Core \cite{garijo_dublin_2013}. Nevertheless provenance data is not often provided on a meta-level \emph{for RDF} data. Instead, entities are introduced that either represent the data (e.g. a \texttt{dcat:CatalogRecord}) or that act as a placeholder for the actual entity for the description from one specific source (e.g. a \texttt{ore:Proxy}). Both approaches are problematic when data from different sources is to be merged due to both approaches must now fit into the same data model.

Even unfavorable is the mixing of data about a resource and about its description. Let's say, we want to model the provenance of a statement that provides a title for an entity $\texttt{<E>}$. In the following example, the information is added directly to the instance of $\texttt{<E>}$.

\begin{verbatim}
<E> rdf:type Entity_E ;
    dct:title "title" ;
    prov:wasDerivedFrom <source> .
\end{verbatim}

But what is derived here? Is it the title of the resource or the resource itself? As all statements describe their subject, it is clearly the resource $\texttt{<E>}$ and not the title or other descriptive data about the resource.

\paragraph{Solving approach:}

Provenance is a prime example where Named Graphs should be used as usually many statements share the same provenance:

\begin{verbatim}
<entity> ex:ID <ID> .

:data {
   <entity>
     ex:data "data" ;
}

:data
   prov:wasDerivedFrom <entity> ;
   ...
\end{verbatim}

\paragraph{Example DBpedia:}
The DBpedia project\footnote{\url{https://www.dbpedia.org/}.} is one of the largest Linked Open Datasets on the Web. It converts Wikipedia articles into RDF. The data is expressed using the DBpedia ontology. The information from which Wikipedia article data was derived is expressed using the \texttt{prov:wasDerivedFrom} predicate.
Additional provenance information is given by further statements such as \texttt{dbo:wikiPageID} giving the ID of the Wikipedia page, \texttt{dbo:wikiPageRevisionID} stating the revision ID or the number of characters of the original article in \texttt{dbo:wikiPageLength}. Indeed, the provenance information is related to the article, but is mixed with the description of the entity the article is about:

\begin{verbatim}
<entity>
   dbp:size "63" ;
   dbp:built "1889" ;
   prov:wasDerivedFrom <article> ;
   dbo:wikiPageID "123" ;
   dct:date "2022-05-21" .
\end{verbatim}

Here the statement level is not sufficient thus multiple statements share the same provenance. A cleaner solution is a model with a meta-level. It can also be argued that the provenance information of the data is only important for a small subset of applications, for example when ensuring the data quality such as the relevancy and timeliness of an article. Here is the example data using Named Graphs:

\begin{verbatim}
:data {
   <entity>
     dbp:size "63" ;
     dbp:built "1889" ;
}

<article> dbo:wikiPageID "123" .

:data
   prov:wasDerivedFrom <article> ;
   dct:date "2022-05-21" ;
\end{verbatim}

\paragraph{Conclusion:}

The meta-level is the best and most suitable way how to model provenance data in RDF: This avoids levels of indirections, i.e. the usage of proxy resources or specific constructs such as the notion of records. The provenance information can be kept separate from the data and may only be queried if needed, which makes the data model lightweight.
While Named Graphs are usually the best fit for provenance data, it has to be noted that RDF-star can also be used -- even additionally -- to provide the provenance for a single statement. This first pattern addresses the most common use case for metadata. With the following two patterns, we would like to extend its use to new applications.

\subsection{Pattern 2: Extending an existing data model}
\label{pattern2}

\paragraph{Problem statement:}
Data models usually evolve over time and changes to the model are inevitable. This poses a problem when these changes are not backwards compatible and break existing applications.
We therefore propose to use the meta-level to extend and improve an existing data model. 

For example, an existing knowledge graph could be enriched to provide additional confidence values. Sometimes an existing data model is wrong, unfitting or simply corrupted resulting in loss of context.

For illustration, we assume we have a simple data model where a relation of two entities $\texttt{<A>}$ and $\texttt{<B>}$ can be expressed:

\begin{verbatim}
<A> :conformsTo <B> .
\end{verbatim}

Now this model is to be enriched with additional information such as a confidence score. To achieve this, however, various approaches are possible without using the meta-level, for example n-ary entities. This would require a remodeling which is a high risk of breaking existing applications, for instance:

\begin{verbatim}
<A> :hasConformanceStatement <C> .
<C> a :ConformanceStatement;
    :conformingTo <B>;
    :confidence 0.8 .
\end{verbatim}

\paragraph{Solving approach:}
We propose a modeling of this new information in the meta-level without touching the existing model to ensure backwards compatibility. The additional information is integrated by using RDF-star.

\begin{verbatim}
<A> :conformsTo <B> .
<< <A> :conformsTo <B> >> :confidence 0.8 .
\end{verbatim}

\paragraph{Example 1 DBpedia:}
\label{dbpedia1}
All data within DBpedia is derived from Wikipedia articles. Among other information, many articles have a thumbnail image related to its article. In DBpedia a thumbnail is attached to an article with the predicate \texttt{dbo:thumbnail}. However, Wikipedia is a Web page where the thumbnail is an HTML image tag. In DBpedia the src-attribute of the tag is converted to the URI describing the thumbnail. Furthermore, an HTML image tag is described by its caption attribute and an alternative text (alt attribute). Both are included in DBpedia via \texttt{dbp:caption} and \texttt{dbp:alt}. However, the relation to which thumbnail a caption or alternative text conforms is gone.\footnote{According to the DBpedia ontology, this should actually be modeled with an n-ary entity, but at least in the current version of DBpedia, the described problem exists.} Due to this simplification the relation is lost. This becomes really disadvantageous if articles have multiple thumbnails where it is impossible to resolve these relationships.
While this obviously should be fixed in the data, it might be that there are applications using the data that rely on the current data representation. With RDF-star, however, we can provide additional information to deliver the relationship information to applications that need it, without changing the asserted statements in the RDF data. One way would be to add the caption to the relationship:

\begin{verbatim}
<entity> dbo:thumbnail <thumbnail> ;
  dbp:caption "Portrait of X" .

<< <entity> dbo:thumbnail <thumbnail> >>
  dbp:caption "Portrait of X" .
\end{verbatim}

This is arguably again not ideal as a caption should probably refer to a thumbnail, not a thumbnail assignment. On the other hand, if the same thumbnail is used in different contexts with different captions, the above solution is good. For the sake of the argument however, let us assume that the caption should actually be assigned to the thumbnail. So the data should look like this:

\begin{verbatim}
<entity> dbo:thumbnail <thumbnail> .
<thumbnail> dbp:caption "Portrait of X" .
\end{verbatim}

To actually fix the data, in this case the subject of the original statement needs to be changed. We could provide a vocabulary for such cases that would be universally understandable by applications supporting it, i.e. the \texttt{replaceSubjectBy}-statement:

\begin{verbatim}
<entity> dbo:thumbnail <thumbnail> ;
  dbp:caption "Portrait of X" .

<< <entity> dbp:caption "Portrait of X" >>
  ex:replaceSubjectBy <thumbnail> .
\end{verbatim}

This means that an application should replace the subject of the original statement (\texttt{<entity>}) with the new subject \texttt{<thumbnail>}. As can be seen here, RDF-star opens many interesting ways to provide a history or change requests for statements. Consider the following example where statements are related to each other:

\begin{verbatim}
<entity> dbo:thumbnail <thumbnail> .
<thumbnail> dbp:caption "Portrait of X" .

<< <thumbnail> dbp:caption "Portrait of X" >>
    ex:replaced << <entity> dbp:caption "Portrait of X" >> .
\end{verbatim}

In this case, the actually wrong statement \texttt{<entity> dbp:caption "Portrait of X".} would only be available as part of the RDF-star triple stating that it has been replaced. Nevertheless, it is still part of the graph and old applications could still use it.

\paragraph{Example 2 GeoNames:}
\label{geonames}
GeoNames\footnote{\url{http://www.geonames.org/}} is a graph for geographical data. It includes alternative names, but not information on historical names. This example shows how historical names of a city, or a place can be added to the current data and extended including more specific information. Currently in GeoNames there is only a distinction of \texttt{gn:name} and \texttt{gn:alternateName}. The alternate name can contain a language tag, but it does not indicate when a name was used in case of historical names.
For example: The German city of Chemnitz used to be called "Karl-Marx-Stadt" between May 10th, 1953 and July 1st, 1990. 

The listing below shows an excerpt of the RDF entry in GeoNames\footnote{\url{https://sws.geonames.org/2940132}.}:

\begin{verbatim}
<https://sws.geonames.org/2940132/> 
  gn:name "Chemnitz" ;
  gn:alternateName "Chemnitz"@de ;
  gn:alternateName "Chemnitz"@en ;
  gn:alternateName "Karl-Marx-Stadt"@de .
    
\end{verbatim}

We could add the missing information about the time span, for example by using the Common Authority File (Gemeinsame Normdatei, GND) of the German National Library, which provides the following data:

\begin{verbatim}
<https://d-nb.info/gnd/2015221-8> 
  gndo:preferredNameForThePlaceOrGeographicName 
    "Karl-Marx-Stadt";
  gndo:dateOfEstablishment "10.05.1953" .
  gndo:dateOfTermination "31.05.1990" ;
\end{verbatim}

We can take this data from the GND and apply it to GeoNames to add the additional data to the entities, such as date of establishment and date of termination:

\begin{verbatim}
<https://sws.geonames.org/2940132/>
  gn:alternateName "Karl-Marx-Stadt"@de .
    
<< <https://sws.geonames.org/2940132/> 
   gn:alternateName "Karl-Marx-Stadt"@de >> 
  ex:valid_from "09.05.1953"^^xsd:date ;
  ex:valid_to "01.06.1990"^^xsd:date .
\end{verbatim}

This way, the data is still provided in a very simple manner that nevertheless is useful for many applications that do not need the additional information, for example for named entity resolution. Nevertheless, the additional information can be provided in a modular way if it is needed.

\paragraph{Conclusion:}
RDF-star is a great fit when additional information about a statement needs to be provided. In particular when a data model is already used in applications it is an adaption to provide more complex data avoiding incompatible changes. By doing so, a simple model retains still simple by providing an additional context in the meta-level. The backwards compatibility is obtained as well (see also Section~\ref{queries}).

\subsection{Pattern 3: Shifting complex relations to the meta-level}\label{pattern3}

\paragraph{Problem statement:} 

Whenever additional data about a relationship between two entities is needed, an additional \emph{n-ary} entity can be created for representation. As entities (subjects and objects in statements) are substantives, this results in graphs with many nominalized relationships, for example:

\begin{verbatim}
<E> :hasSubject <SubjectAssignment1> .
<SubjectAssignment1>
    :hasHeading "Data Modeling" ;
    :fromVocabulary <TopicsVocabulary> .                
\end{verbatim}

Here, a complex entity of the class \texttt{SubjectAssignment} is used to represent the subject of entity \texttt{<E>}. The problem is that this structure feels cumbersome for many applications that might only be interested in the subject heading and that do not care about the vocabulary the subject heading is coming from. Even if uncontrolled subject headings (free tags) are used, the intermediate subject assignment is still required as the data model is created this way.

\paragraph{Solving approach:} 
Similar to pattern 2, we propose to use a meta-level, but this time from the beginning, so that the additional and perhaps even optional information can be pushed to the meta-level:
\begin{verbatim}
<E> dc:subject "Data Modeling" .
<< <E> dc:subject "Data Modeling" >> 
    :fromVocabulary <TopicsVocabulary> .
\end{verbatim}

\paragraph{Example DCAT:}
\label{experiments_dcat}
In this example we demonstrate how such n-ary entities can be avoided directly at the time of designing ontologies by shifting entities to the meta-level.
DCAT is an ontology enabling publishers to describe datasets and its properties. The datasets can be listed in a \texttt{dcat:Catalog} entity. The \texttt{dcat:Dataset} entity holds meta information of the actual data, which is linked to as a \texttt{dcat:Distribution} entity. When listing a dataset in the catalog, the optional entity \texttt{dcat:CatalogRecord} may be used to express metadata about the listing such as the issue date.

As of the current DCAT 2 ontology specification\footnote{\url{https://www.w3.org/TR/vocab-dcat-2/}.}, the \texttt{dcat:CatalogRecord} is provided for the purpose of adding additional metadata for the description of the listing of a resource, e.g., datasets in the catalog. The following example lists a dataset in a catalog. To add metadata such as the issued date or the title, the \texttt{CatalogRecord} entity is applied providing these information.

\newpage

\begin{verbatim}
ex:catalog dcat:record ex:catalogRecord .

ex:dataset a dcat:Dataset .

ex:catalogRecord a dcat:CatalogRecord ;
  dct:issued "05.04.2022" ;
  dct:title "record title" ;
  dct:description "record description" ;
  foaf:primaryTopic ex:dataset .
\end{verbatim}

However, the \texttt{dcat:CatalogRecord} entity adds metadata only to the actual listing of the catalog and the dataset. This can be shortened by rewriting it in the following RDF-star syntax to shift the \texttt{dcat:CatalogRecord} entity to the meta-level. The attributes of this entity can be used to describe the provenance of the relation directly: 

\begin{verbatim}
ex:catalog dcat:dataset ex:dataset .

<< ex:catalog dcat:dataset ex:dataset >>
  a dcat:CatalogRecord ;
  dct:issued "05.04.2022" ;
  dct:title "record title" ;
  dct:description "record description" .
\end{verbatim}

This information can be queried easily from the meta-level using SPARQL-star. Here we give an example on how to query the issued date of the CatalogRecord denoted before:

\begin{verbatim}
SELECT ?date WHERE {
  << ex:catalog dcat:dataset ex:dataset >> dct:issued ?date .
}
\end{verbatim}

As RDF data can easily be split in several junks of statements, the metadata could also be separated from the core data, containing only the RDF-star statements.

\paragraph{Conclusion:} On the one hand, complexity can be shifted to the meta-level and the data model is improved towards simpleness. In the course of this, the data may only be queried when needed reducing the size of query results.
On the other hand, the simple data model might seem to be oversimplified to a data modeler. If the vocabulary scheme belongs to the subject heading and is not optional, it might seem arbitrary to put this information to the meta-level. And if the majority of applications need the complete data, the querying and using of the meta-level can feel more cumbersome than a more complex data model.
To conclude, there are situations where the data modeled in an entity really should be on the same level as the rest of the actual data. In other situations, using the meta-level might just be the ideal solution where a simple model can be provided for simple applications and additional information is available if needed.

\section{Querying and constructing the meta-level with SPARQL-star}
\label{queries}

In the last section we have shown how data using the meta-level in its data model could look like. For a broad adoption of the proposed patterns, it is important that working with the data is effortless and data on the meta-level can easily be found and processed when needed.
For this, two aspects should be considered: 
\begin{enumerate}
    \item How can an application find out if there is data on the meta-level?
    \item How can data using the meta-level be transformed to other data models that potentially do not use the meta-level?
\end{enumerate}

\paragraph{How can an application find out if there is data on the meta-level?}

For this question, it is important to adjust some expectations that an application might have when dealing with RDF and Linked Data. A modern RDF application that supports data on the meta-level is required 
\begin{enumerate}
    \item  to support Named Graphs and RDF-star as well as at least TriG-star\footnote{\url{https://w3c.github.io/rdf-star/cg-spec/2021-04-13.html\#trig-star}} as standard format for any responses from a Linked Data server as well as for data dumps,
    \item to expect Named Graphs to be used for the actual data, so the internal organisation of data, e.g. for provenance tracking, must support graph hierarchies, such as a provenance context~\cite{eckert_provenance_2013},
    \item to check for the existence of a meta-level in form of Named Graphs and/or RDF-star triples.\footnote{\url{https://w3c.github.io/rdf-star/cg-spec/2021-04-13.html\#dfn-triple}}
\end{enumerate}

The following SPARQL-star query returns all RDF-star triples:

\begin{verbatim}
SELECT DISTINCT ?s WHERE {
  ?s ?p ?o .
  FILTER( isTRIPLE(?s) )
}
\end{verbatim}

The function \texttt{isTRIPLE} is new in SPARQL-star and returns \texttt{TRUE} if the parameter is an RDF-star triple. Together with a query for the existence of Named Graphs, this enables an application to check if a meta-level is available.

\paragraph{How can data using the meta-level be transformed to other data models that potentially do not use the meta-level?}

The ability to transform data from one data model to another by means of \texttt{CONSTRUCT} queries is one of the most powerful features of RDF. This is not restricted by the use of meta-level data, as the following queries demonstrate:

In Pattern~3 (Section~\ref{pattern3}) we have shown how to shift the meta information encapsulated in an n-ary entity to the meta-level. Using SPARQL-star and CONSTRUCT queries, it is possible to rebuild the former data model, e.g. for the \texttt{dcat:CatalogRecord}. To extract the subject or object of a triple (annotated with RDF-star), the SPARQL-star functions \texttt{SUBJECT} and \texttt{OBJECT} are used:

\begin{verbatim}
CONSTRUCT {
  ?catalog :record ?record .
  ?record :primaryTopic ?dataset .
  ?record ?pred ?obj .
} WHERE {
  ?star ?pred ?obj 
  FILTER(isTRIPLE(?star)) . 
  BIND(SUBJECT(?star) as ?catalog) .
  BIND(OBJECT(?star) as ?dataset) .
  BIND(IRI(CONCAT(STR(?dataset), "/record")) as ?record) .
}
\end{verbatim}

Here, string functions are used to coin an IRI for the new n-ary entity \texttt{?record}, by appending \texttt{/record} to the IRI of \texttt{?dataset}. This is to avoid a blank node, but it depends on the application if and how this should done.

The other way round is also possible, i.e., to create a simple data model with meta-level data from a complex data model:

\begin{verbatim}
CONSTRUCT {
  ?star ?pred ?obj .
} WHERE {
  ?catalog :record ?record .
  ?record :primaryTopic ?dataset .
  ?record ?pred ?obj .
  BIND(TRIPLE(?catalog, :dataset, ?dataset) as ?star).
}
\end{verbatim}

This time, the SPARQL-star function \texttt{TRIPLE} is used to create an RDF-star triple from a subject, a predicate, and an object.

\section{Discussion}

With this paper, we ask the following question: can and should meta-level concepts be used for data modeling in RDF? Besides Named Graphs and RDF-star, several concepts have been introduced to construct a meta-level as well. However, they all have weaknesses regarding e.g., feasibility or performance. Named Graphs have been around for a long time now and are well-supported in SPARQL. With RDF-star, we now see a serious contender for an RDF extension that allows statements about statements in a concise and straight-forward way, also with a SPARQL extension. So yes, we can use it for data modeling. But should we?

We identified three patterns that are suitable for a more detailed discussion of the potential benefits of meta-level modeling: (1) using the meta-level for actual metadata, i.e. data about RDF data like provenance data; (2) using the meta-level for the extension of existing data models; and (3) using the meta-level from the get-go to get simpler and more modular data models.

The first pattern is arguably the most trivial one, but addresses the reasons why Named Graphs and RDF-star have been developed in the first place. While at least Named Graphs exist for quite some while now, we still see only rarely data models that make use of them. Instead, mostly workarounds within vanilla RDF are used to artificially create a meta-level when it is needed. This should change and data about RDF data should be where it belongs: in the meta-level.

With the second pattern, we showed that existing simple data models can be extended by additional meta statements, with two main scenarios in mind: First, additional data can be added to provide more complex information to applications that need it. Second, this can be used to provide corrections and missing information to a data model without compromising existing applications.
Adding the additional information in the meta-level does not only preserve backwards compatibility of the data model. It also keeps the core model simple and allows a modular distribution of additional data for the applications that need it. 
This observation leads to the third pattern where we proposed to use the meta-level to create simple and modular data models from the beginning.

We have further shown in several examples from real world data (including DCAT, DBpedia, and GeoNames), how the application of these patterns could look like and how they would affect the provided data. 
And finally, we have shown that working with the meta-level is straight-forward, as long as applications expect a meta-level and support Named Graphs and RDF-star, with TriG-star as preferred serialization. Particularly also the transformation of data into and out of the meta-level using \texttt{CONSTRUCT} queries is possible.

For the question, if meta-level modeling should be used, we conclude with the following thoughts:

\paragraph{Is the provision and management of meta-level data possible?} This depends on the technology stack that is used. Named Graphs are usually not a problem for triple stores but might lead to problems when the meta-level is to be preserved in other systems. For data serializations, a compatible file format such as TRIG must be used. File-based representations of graphs (i.e., one RDF file per graph) are also possible, but require additional infrastructure and/or tooling to make sure all meta information is properly preserved. RDF-star is very new, but at least a transparent serialization via RDF reification is possible.

\paragraph{Is there a clear decision for Named Graphs and/or RDF-star?} As stated in the paper, it depends on the use case and sometimes even on the data. It might be worthwhile to just use one of the mechanisms to reduce complexity within a project. Alternatively, both can be used, but for different parts of the data model, for instance Named Graphs to provide provenance information and RDF-star to provide additional information on entity links. And finally, this could be left to the implementation and applications would have to expect such information being provided with either of the mechanisms.

\paragraph{Can we still "follow the nose"?} One of the main advantages of RDF is that it is easy to get data about a resource and use subsequent queries to find out more or even just to check if there are any more information. This is in our opinion the main reason why meta-level information is not used as often as it should. Applications often expect Turtle or even still RDF/XML as RDF serializations. They might not be able to deal with TriG. With SPARQL, Named Graphs can be queried. However, the application needs to know that there is useful information attached to a Named Graph. This is similar for RDF-star. The application needs to ask (or understand from the serialization) if there are additional statements about a statement. On the other hand, checking for the existence of Named Graphs or RDF-star triples is possible in SPARQL and SPARQL-star. So this is only a matter of getting used to it.

\section{Conclusion and Future Work}

Using the meta-level actively for data modeling opens interesting new perspectives on the task. In this paper we formulated three abstract patterns on how and when to use the meta-level in the modeling process actively. Therefore we gave several examples on real world applications.

Usually, the data modeler has to decide if the model should be optimized for data consumption, i.e. as easy to digest as possible and only as complex as necessary. However, more often than not, it is rather created with the most complex possible data in mind. Often data that does not even exist (yet) in reality - but certainly would be great to have at some point so the data model should better be ready. Meta-level modeling actually provides a means to define a simple core data model - particularly with simple, direct relations. Additional data can be provided in a modular way, with additional RDF data that can contain additional statements further describing resources.

For future work, we aim at using the meta-level in our own projects on production level to gain further experience with it and to support a wider adoption of Named Graphs and RDF-star in public data models. We are interested in the potential of meta-level modeling for the representation of changes in data to support data transparency and better provenance tracking. Currently we are exploring the creation of an ontology to capture the abstract entities appearing in meta-modeling, based on the patterns we identified so far and potentially more patterns that will arise with further applications.

\section*{Acknowledgements}
This research was partially supported by the Volkswagen Foundation (Project: Consequences of Artificial Intelligence on Urban Societies, Grant 98555) and the German Research Foundation (Project: Specialized Subject Service for Jewish Studies, Grant 286004564). We thank Magnus Pfeffer for his valuable feedback.

\bibliographystyle{splncs03}
\bibliography{literature.bib}

\begin{thebibliography}{10}
\providecommand{\url}[1]{\texttt{#1}}
\providecommand{\urlprefix}{URL }

\bibitem{noauthor_resource_nodate}
Resource {Description} {Framework} ({RDF}): {Concepts} and {Abstract} {Syntax},
  \url{https://www.w3.org/TR/rdf-concepts/}

\bibitem{bucur_unified_2020}
Bucur, C.I., Kuhn, T., Ceolin, D.: A {Unified} {Nanopublication} {Model} for
  {Effective} and {User}-{Friendly} {Access} to the {Elements} of {Scientific}
  {Publishing}. In: Knowledge {Engineering} and {Knowledge} {Management}. pp.
  104--119. Springer (2020), \url{https://arxiv.org/abs/2006.06348}

\bibitem{carroll_named_2005}
Carroll, J.J., Bizer, C., Hayes, P., Stickler, P.: Named graphs, provenance and
  trust. In: Proceedings of the 14th international conference on {World} {Wide}
  {Web} - {WWW} '05. pp. 613--622. ACM Press, Chiba, Japan (2005)

\bibitem{dodds_follow_2012}
Dodds, L., Davis, I.: Follow {Your} {Nose}. In: Linked {Data} {Patterns} - {A}
  pattern catalogue for modelling, publishing, and consuming {Linked} {Data}
  (2012), \url{https://patterns.dataincubator.org/book/follow-your-nose.html}

\bibitem{eckert_provenance_2013}
Eckert, K.: Provenance and {Annotations} for {Linked} {Data}. In: Linking to
  the {Future}: 2013 {Proceedings} of the {International} {Conference} on
  {Dublin} {Core} and {Metadata} {Applications}. pp. 9--18 (2013)

\bibitem{garijo_dublin_2013}
Garijo, D., Eckert, K.: Dublin {Core} to {PROV} {Mapping} (2013),
  \url{https://www.w3.org/TR/prov-dc/}

\bibitem{hartig_foundations_2017}
Hartig, O.: Foundations of {RDF}* and {SPARQL}*. In: Proceedings of the 11th
  {Alberto} {Mendelzon} {International} {Workshop} on {Foundations} of {Data}
  {Management} and the {Web} 2017. {CEUR} {Workshop} {Proceedings}, vol. 1912,
  pp. 1--11 (2017)

\bibitem{heery_application_2000}
Heery, R., Patel, M.: Application {Profiles}: {Mixing} and {Matching}
  {Metadata} {Schemas}. Ariadne (25) (2000),
  \url{http://www.ariadne.ac.uk/issue/25/app-profiles/}

\bibitem{isaac_antoine_europeana_2013}
Isaac, A.: Europeana {Data} {Model} {Primer} (2013),
  \url{https://pro.europeana.eu/files/Europeana_Professional/Share_your_data/Technical_requirements/EDM_Documentation/EDM_Primer_130714.pdf}

\bibitem{lagoze_ore_2008}
Lagoze, C., Van~de Sompel, H., Johnston, P., Nelson, M., Sanderson, R., Warner,
  S.: {ORE} {User} {Guide} - {Primer} (2008),
  \url{http://openarchives.org/ore/1.0/primer}

\bibitem{nguyen_dont_2014}
Nguyen, V., Bodenreider, O., Sheth, A.: Don’t {Like} {RDF} {Reification}?
  {Making} {Statements} about {Statements} {Using} {Singleton} {Property}.
  Proceedings of the ... International World-Wide Web Conference. International
  WWW Conference  2014,  759--770 (Apr 2014)

\bibitem{olaf_hartig_rdf-star_2021}
{Olaf Hartig}, {Pierre-Antoine Champin}, Kellogg, G., Seaborn, A.: {RDF}-star
  and {SPARQL}-star (2021), \url{https://www.w3.org/2021/12/rdf-star.html}

\bibitem{orlandi_benchmarking_2021}
Orlandi, F., Graux, D., O'Sullivan, D.: Benchmarking {RDF} {Metadata}
  {Representations}: {Reification}, {Singleton} {Property} and {RDF}. In: 2021
  {IEEE} 15th {International} {Conference} on {Semantic} {Computing} ({ICSC}).
  pp. 233--240 (Jan 2021), iSSN: 2325-6516

\bibitem{orlandi_modelling_2011}
Orlandi, F., Passant, A.: Modelling {Provenance} of {DBpedia} {Resources}
  {Using} {Wikipedia} {Contributions}. Journal of Web Semantics  Volume 9(Issue
  2),  149--164 (2011)

\bibitem{patrick_j_hayes_rdf_2014}
Patrick J.~Hayes, P.F.P.S.: {RDF} 1.1 {Semantics} (2014),
  \url{https://www.w3.org/TR/rdf11-mt/}

\bibitem{perez_systematic_2018}
Pérez, B., Rubio, J., Sáenz-Adán, C.: A systematic review of provenance
  systems. Knowledge and Information Systems  57(3),  495--543 (Dec 2018)

\bibitem{sikos_provenance-aware_2020}
Sikos, L.F., Philp, D.: Provenance-{Aware} {Knowledge} {Representation}: {A}
  {Survey} of {Data} {Models} and {Contextualized} {Knowledge} {Graphs}. Data
  Science and Engineering  5(3),  293--316 (Sep 2020),
  \url{https://link.springer.com/10.1007/s41019-020-00118-0}

\bibitem{svensson_indicating_2021}
Svensson, L.G., Verborgh, R., Sompel, H.V.d.: Indicating, {Discovering},
  {Negotiating}, and {Writing} {Profiled} {Representations}. Internet {Draft}
  draft-svensson-profiled-representations-01, Internet Engineering Task Force
  (Mar 2021),
  \url{https://datatracker.ietf.org/doc/draft-svensson-profiled-representations-01}

\end{thebibliography}

\renewcommand{\indexname}{Author Index}
\printautindex

\end{document}